\documentclass[a4paper,11pt]{article}
\pdfoutput=1
\usepackage{color}
\usepackage{jinstpub}
\usepackage{lineno}
\usepackage{float}
\usepackage{graphicx}
\usepackage{epstopdf}
\usepackage{graphicx}
\usepackage{subfigure}
\usepackage{url}
\usepackage{multirow}
\usepackage{listings}
\usepackage{color}
\usepackage{pythonhighlight}

\title{Slow Control System for PandaX-III experiment}
\author[a]{Xiyu Yan,} 
\author[b,1]{Xun Chen,}
\author[c]{Yu Chen,}
\author[d]{Bo Dai,}
\author[b]{Heng Lin,}
\author[c]{Tao Li,}
\author[b]{Ke Han,}
\author[b]{Kaixiang Ni,}
\author[d]{Fusang Wang,}
\author[b,d,1]{Shaobo Wang,\note{Corresponding authors.}}
\author[a]{Qibin Zheng,}
\author[b]{Xinning Zeng}

\affiliation[a]{Institute of Biomedical Engineering and Terahertz Technology Innovation Research Institute, University of Shanghai for Science and Technology, Shanghai 200093, China}
\affiliation[b]{INPAC; Shanghai Laboratory for Particle Physics and Cosmology; 
 Key Laboratory for Particle Astrophysics and Cosmology (MOE), \\
School of Physics and Astronomy, Shanghai Jiao Tong University, Shanghai {\rm 200240}, China}
\affiliation[c]{School of Physics, Sun Yat-sen University, Guangzhou, 510275, China}
\affiliation[d]{SPEIT~(SJTU-ParisTech Elite Institute of Technology), Shanghai Jiao Tong University, Shanghai 200240, China}

\emailAdd{shaobo.wang@sjtu.edu.cn;chenxun@sjtu.edu.cn}

\abstract{The PandaX-III experiment uses high pressure gaseous time projection chamber to search for the neutrinoless double beta decay of $^{136}$Xe. A modular slow control system~(SCS) has been designed to monitor all the critical parameters of the experiment.
It ensures the proper operation of the experiment as well as to provide necessary information for data corrections. The main subsystems of the experiment will be constantly monitored by the data collection module of the SCS,
which collects data from sensors and sends them to the centralized database.
When an alarm message is generated by the anomaly detection module,
it will be sent to an alert website and related on-call experts immediately.
A demonstrator of the SCS has been built for the PandaX-III prototype detector. 
The long-term test of it provided valuable experience for the final design of the SCS for PandaX-III.
}

\arxivnumber{2012.13275}

\begin{document}
\maketitle
\flushbottom

\section{Introduction}

The purpose of the planned PandaX-III experiment~\cite{Chen:2016qcd}
is to search for the neutrinoless double beta decay~(NLDBD) of
$^{136}$Xe isotope.  It employs a high pressure gaseous time
projection chamber~(TPC)~\cite{Wang:2020owr} locating in the CJPL-II,
with 140~kg 90\% $^{136}$Xe-enriched at the pressure of 10~bar.  The
TPC collects only ionized electron signals with 52 Micromegas
detectors~\cite{Giomataris:1998rc}.  To achieve the required energy
resolution of 3\% full width at half maximum~(FWHM), about 1\% of
TMA~(Trimethylamine) will be mixed with xenon to enhance the
ionization.  A gas handling system is designed to pump the pipelines
and the vessel, as well as to do the circulation to ensure the purity
of the working gas.  An internal calibration system is to inject
$^{220}$Rn and $^{83m}$Kr into the detector through the gas
circulation loop~\cite{Wang:2020csx}.  Many parameters need to be
monitored during the gas handling.  In order to ensure the safe
operation of the detector, the high voltage and the current of all the
Micromegas need to be monitored.  The critical parameters of the other
subsystems are also important for the running of the experiment and
data analysis. In addition, abnormal states of the detector and the
subsystems require a timely alarm to be handled in time.

We designed a slow control system~(SCS) for the PandaX-III experiment,
borrowed the modular idea from that in the PandaX-4T
experiment~\cite{Ji:2019cwn}. A pool of clients collect data from
different subsystems of the experiment and send them to the
centralized database, which stores the value and timestamp of data.
The evolution of data can be viewed through the web-based interface
and the abnormal events can be detected.  In comparison with that
reported in Ref.~\cite{Ji:2019cwn}, significant improvements on the
data collection module have been made, and a new alert dispatching
module has been created in this system, which will send alter messages
to an alarm website and related people.

In this report, we present the design of the PandaX-III SCS and the
demonstrator in this report. The design of the system, together with
the improvements, are presented in Section~\ref{System design}.  The
integration of system is described in Section~\ref{intergration}.  The
set up and operation of the demonstrator are introduced in
Section~\ref{demonstrator}.  In the final section,
Section~\ref{conclusion}, we give the conclusion.

\section{System design}
\label{System design}
The SCS is composed of five modules, corresponding to the function of
data storage, collection, visualization, anomaly detection and alarm
dispatching. Each module is loosely coupled with others and can work
standalone, simplifying the test and deployment of the system, and
improving the fault tolerance. The data storage module provides a database for
the whole system. It stores the history data of the monitoring
parameters. The data collection module collects data from the
subsystems of the experiment and sends them to the data storage module.
The visualization module is used to display the evolution of the
parameters over time. The anomaly detection module is responsible for
the detection of unexpected change of the data and generates alert
reports. The alert messages will be delivered to the alarm system and
on-duty experts by the alert dispatching module.

\subsection{Data storage module}
\label{storage}
The open-source database software of InfluxDB (version 1.7.9)~\cite{InfluxDB},
provided by InfluxData Inc., is used as the data storage module. It is
optimized to store the time series data, expressed as {\em
  measurements} within the database. Each {\em measurement} is in the
form of key-value pairs, where the key is the name and the value is the
actual data. Arbitrary keys can be used during the insertion of a
measurement. A set of tags, in the form of key-value pairs, can be
added to the {\em measurement} to provide additional description of
the data. Indices are built by InfluxDB automatically based on the
tags to enable fast data selection.

InfluxDB provides application programming interfaces (APIs) over the
hypertext transfer protocol (HTTP) for data insertion, querying and
deletion, with the standard GET, POST, DELETE actions. This feature
simplifies the development of related client programs. Once an
instance of InfluxDB is started, a local port is exposed for the
related access and no more works required, making the deployment of
the module quite simple.

\subsection{Data collection module}
\label{collection}
The data collection module contains many data collection clients,
where each works independently and relies on specified hardware and
software. The purpose of the clients is to acquire and decode the
parameter values from the subsystems, and deliver them to the
database.

The clients take different forms. For example, a Windows desktop
program to control the high voltage (HV), developed with the \emph{C\#}
language~\cite{c-sharp}, is extended. So it can send the required data
to the InfluxDB database directly. One of the most important form of
the client is a Python-based~\cite{Python} program running on Linux
boxes. The program is inherited from that developed in
PandaX-4T~\cite{Ji:2019cwn}, with four significant improvements:
\begin{enumerate}
\item The script is running in the ``on-shot'' mode~(Appendix~\ref{On-Shot}) currently, which
  means they will read the data and send them to the database only
  once, then exit immediately. This mode simplified the management of
  multiple-processes and schedules, leading to a more robust
  system. To control the running of the script, we use the ``Timer''
  feature provided by the system service manager~\cite{systemd}. In
  addition to the ``Service'' unit file, another ``Timer'' file is
  required to accomplish this task. Example codes of the control files
  for one client is given in Appendix~\ref{Timer}.
\item For each readout values, the corresponding timestamp is
  determined by the client program currently, while in the previous
  implementation, it is generated automatically by the InfluxDB when
  the data is being inserted. Such an update results in more accurate
  time information, because the lagging from network connections
  cannot be avoid.
\item The client program will write the readout values to a local file as a temporary data storage
  when InfluxDB cannot be reached for some reasons, and then fill all
  these values to the database when it is back, avoiding the loss of
  slow control information.
  The codes and an example of test are given in Appendix~\ref{back up}.
\item Additional types of data sources are supported by the
  program. The program can read data from text files in the format of
  comma-separated values(CSV), which are generated by third-party
  programs to control some of the subsystem periodically. The air
  conditional control program of the clean room currently being built is an
  example.
\end{enumerate}

The low cost Raspberry Pi 4 Model B (Pi) single-board
computers~\cite{RaspberryPi:4B} are used to execute the client program
to take data via the RS-485~\cite{RS485} or RS-232~\cite{RS232}
protocols. During the test, more than ten external devices can be
connected to the same Pi, with the help of serial to USB cables,
RS-485 hubs and USB hubs. Several instances of the client program are
running on the server host the database.

In general, a client program consumes about 20~MB of memory at
running. Even the Pi has enough resource for many client programs to
work at the same time. We estimate that only five Pi and one x86
server is enough for handling the hundreds of parameters in the SCS of
PandaX-III.

\subsection{Data visualization and anomaly detection modules}

\begin{figure}[htbp]
  \renewcommand{\figurename}{Fig.}
  \centering
  \includegraphics[scale=0.21]{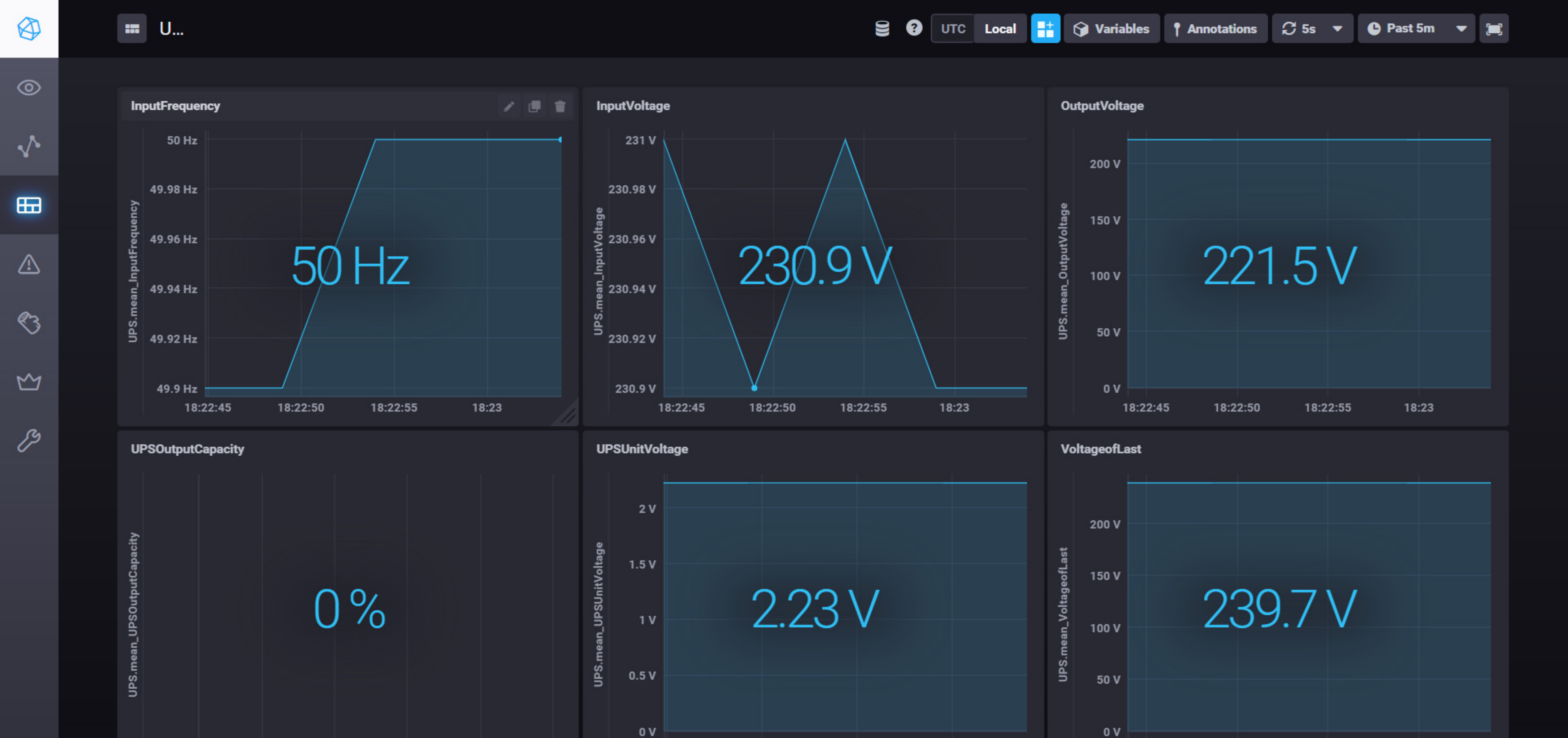}
  \caption{Example of the custom dashboard. The dash board displays
    the time curve and current value of parameters of the UPS, such as
    the input voltage and the frequency.}
  \label{dashboard}
\end{figure}

Chronograf (version 1.7.16), another open source software provided by
InfluxData~\cite{InfluxDB}, is used as the data visualization
module. It provides a web-based user interface to visit the data
stored in InfluxDB. One can use the function of data explorer to
investigate any {\em measurement} directly by constructing data
queries by clicking mouse. The dashboard function provides an interface
for the visualization of selected {\em measurements} in a predefined
style. The user can create many dashboards, each for a group of
interested {\em measurements}. The style of a dashboard can be
customized easily. The queries constructed by the data explorer can be
sent directly to any dashboards. This is convenient when new {\em
  measurements} are added to the database. An example of the
dashboard is shown in Fig.~\ref{dashboard}. User can change the time range
to display the data.

InfluxData also provides Kapacitor (version 1.5.3)~\cite{InfluxDB},
which is a data processing program integrated tightly with InfluxDB
and Chronograf, and is used as the anomaly detection module of the
SCS. The software can process incoming data in real time, and find the
abnormalities according to rules defined by users using the Tickscript
language. For example, a user can set the upper and lower limit for a
{\em measurement}, then Kapacitor will generate an alert message when
the value is out of range. The rules can be created directly in the
``Alerting'' interface provided by Chronograf. The content of the
alert message can be customized. Kapacitor will also deliver the alert
message to predefined event handlers, such as an email address or some
web APIs operated by third-party Internet service providers. We use
the ``Post'' event handler to send the alert messages to the alert
dispatching module.
 
\subsection{Alert dispatching module}
\begin{figure}[htbp]
  \renewcommand{\figurename}{Fig.}
  \centering
  \subfigure[Normal status]{
    \label{fig:subfig:a}
    \includegraphics[scale=0.11]{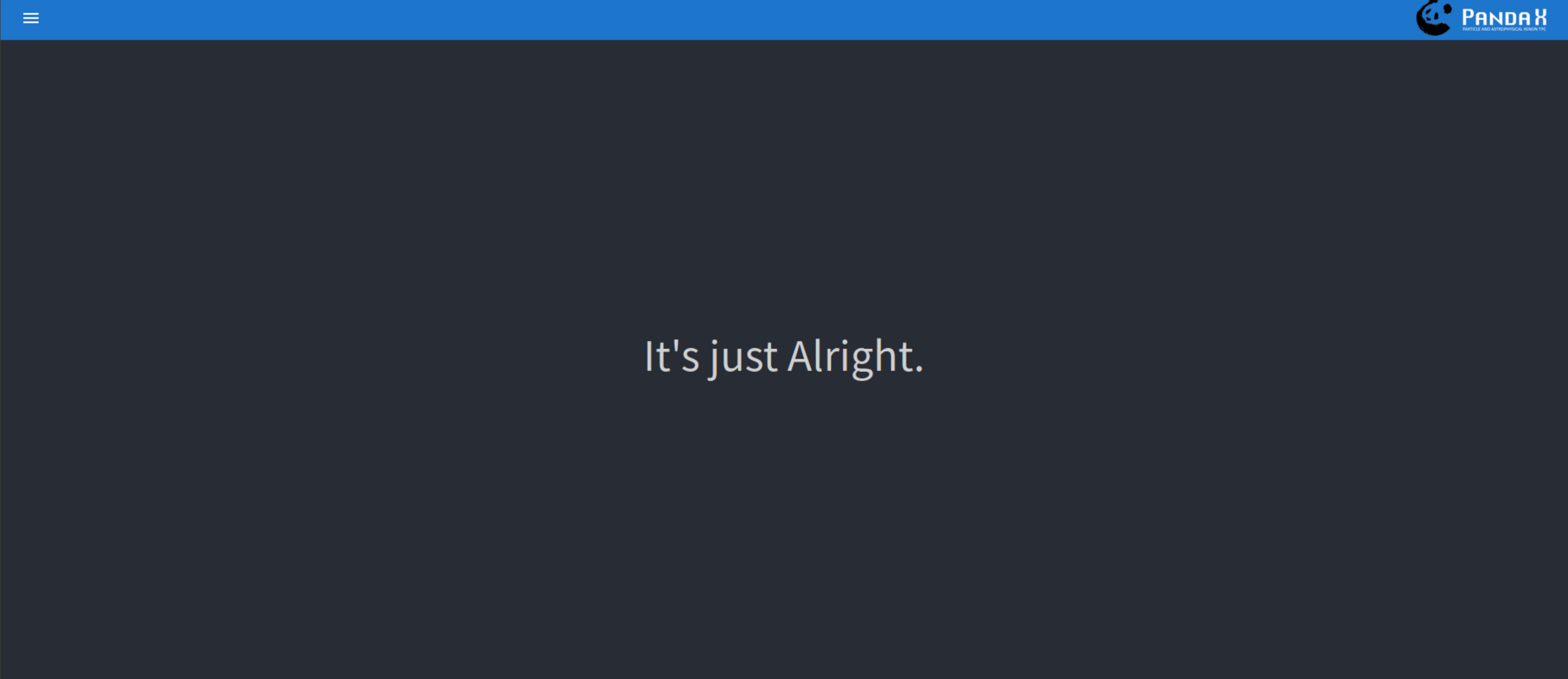}	
  }
  \subfigure[Alert status]{
    \label{fig:subfig:b}
    \includegraphics[scale=0.1]{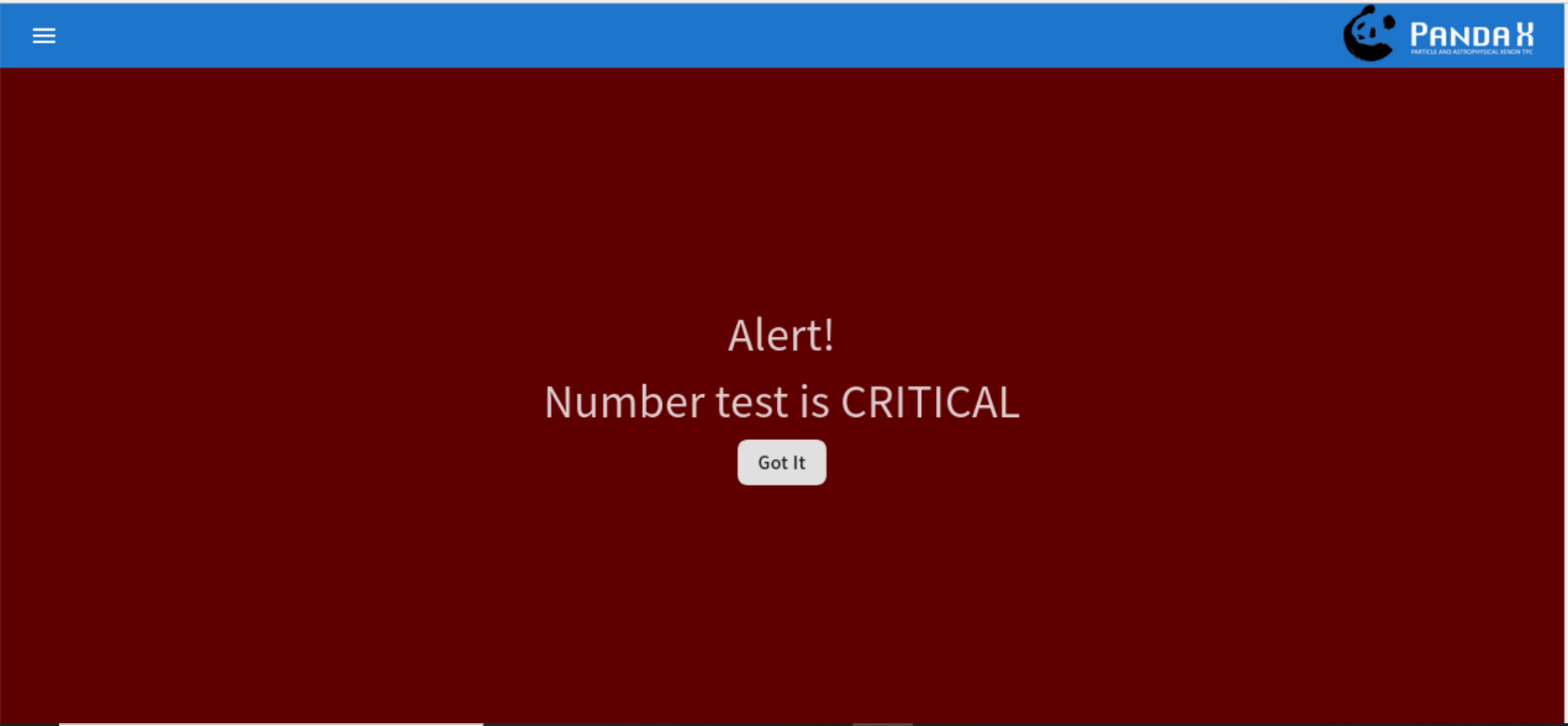}
  }
  \subfigure[Alert history]{
    \label{fig:subfig:c}
    \includegraphics[scale=0.214]{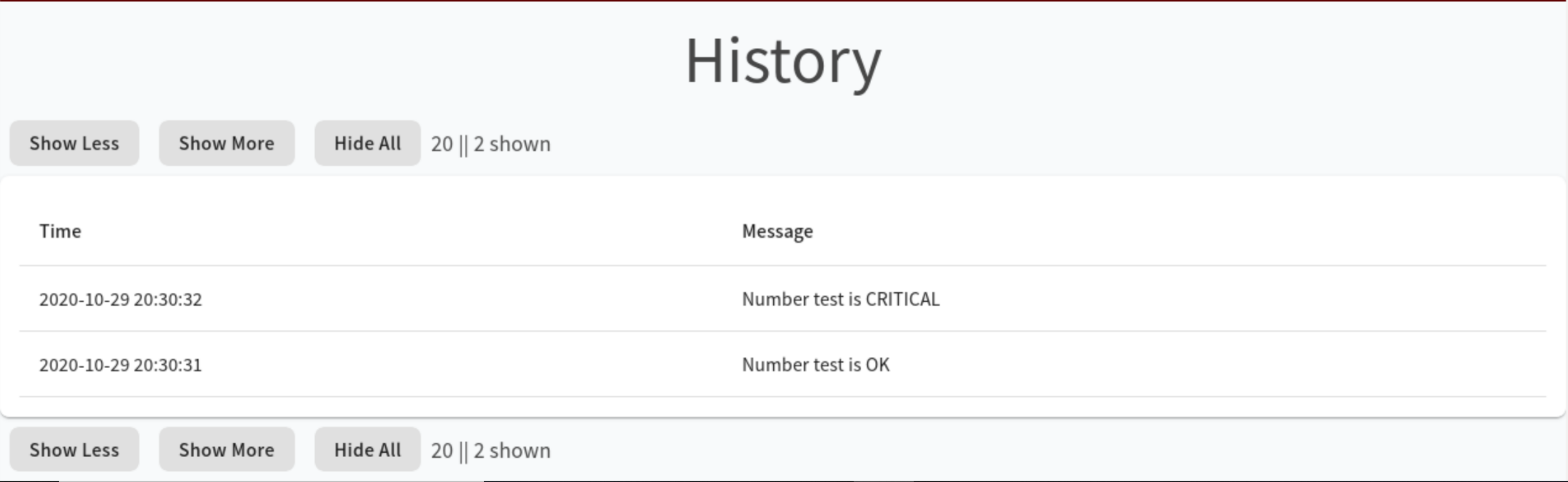}
  }
  \caption{Interface of alert page. The normal status of page with black background is
    shown in (a). An alert message with red background
    is displayed on the center of the page (b). The historical alert
    messages can be displayed within a certain period of time as shown in (c).}
  \label{alert page}
\end{figure}

We created a web application as the alert dispatching module. The
server-side is developed with the Python language, using the Falcon
web framework~\cite{falcon} and the Gunicorn http
server~\cite{Gunicorn}. The server-side can accept the alert message
produced by the Kapacitor. The message will be displayed on the web
page and be sent to the Enterprise WeChat service provided by
Tencent~\cite{WeChat}. The browser-side of the application is
developed with ReactJS~\cite{reactjs}. It can display the current
alert message and the alert history (see Fig.~\ref{alert page}). Alarm
sound will be played by the browser when an alert message arrives
until it is acknowledged by the onsite people. Besides, remote experts
will receive the alert message from the Enterprise WeChat from their
smart phones. The list of remote expert names can be configured.

The design of the alert dispatching module enables the onsite people
and remote experts to act in time at different scenarios of
emergency. For example, when the sparking of the Micromegas detector
happens, onsite people are expected to be aware of the situation
immediately, and the loud alarm sound can be very convenient. The
remote experts may provide necessary help to the onsite people,
according to the alert information received.

\section{Integration with the PandaX-III facility}
\label{intergration} 
\begin{figure}
  \renewcommand{\figurename}{Fig.}
  \centering
  \includegraphics[width=1\textwidth]{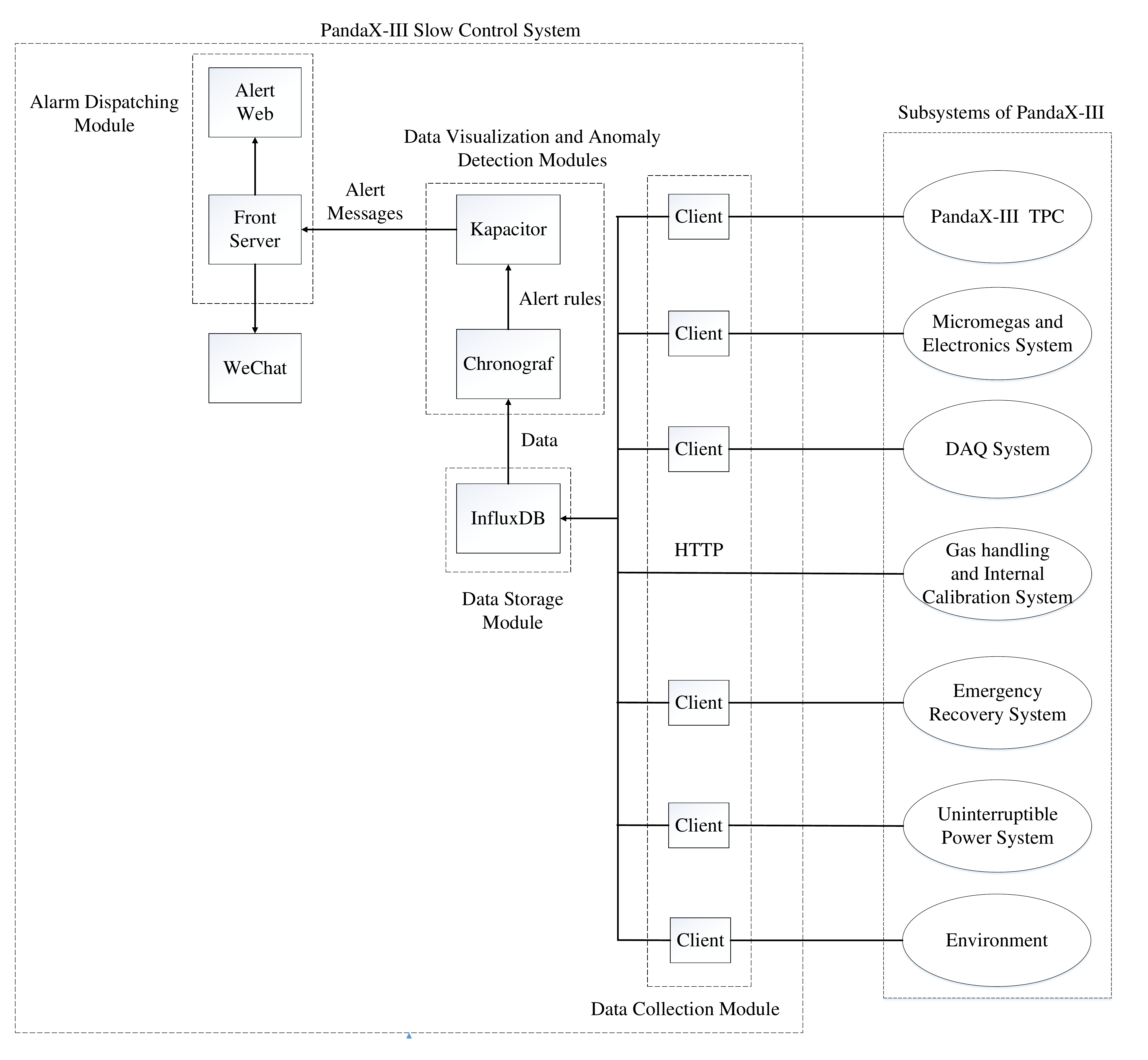}
  \caption{Schematic view of the integration of the PandaX-III SCS and
    the facility.}
  \label{overview}
\end{figure}

The PandaX-III facility consists of six subsystems. The high pressure
xenon gas detector is the core component of PandaX-III. The
electronics, together with the Micromegas detector, are used to
collect and amplify the electron signals. The DAQ system is
responsible for the data taking. The gas handling system works for the
gas circulation as well as the internal calibration. Xenon gas will be
recovered at the time of emergency by the recovery system. Their work
conditions are required to be monitored continuously.  The
uninterruptible power system (UPS) provides backup power supply to the
experimental facility as well as the SCS in the occasions of power
failure.  Besides, the laboratory environment is also to be
monitored. The integration of the SCS to these subsystems is shown in
Fig.~\ref{overview}.

The data of the detector will be collected by the client programs
running on the slow control server due to the short distance from the
server rack to the detector. The most important parameters of the
detector are the HV and the current of the
cathode. They are indicators of the operation state of the
experiment. The HV of cathode will be provided by the Matsusada AU HV
power supplies~\cite{Matsusada:AU}, and can be readout via the serial
to USB cable with RS232 protocol. Besides, a pressure sensor (model
PRC-905\cite{prc}) produced by the HangZhou RunChen Technology Inc.,
will be placed inside the chamber to record the inner pressure of the
TPC.  The output voltage of the pressure sensor (ranging from
1$\to$5~V) will be converted to digital signals, which will be
collected by an instance of the client program via serial cable.  A
thermocouple will be used to measure the temperature of the TPC.  The
digital values will be collected and monitored in the same way.
	 
The HV of the Micromegas detector and the input voltage of the
front-end card (FEC)~\cite{Zhu:2018oxf} are crucial for the collection
and readout of the signals. They will be both provided by the
CAEN SY2527~\cite{CAEN} multi-channel power supply system. Two slots of
the system, with 16 output channels in total, are used for the HV of
the Micromegas, while one slot with 8 channels is used for the power
supply of the FEC. The values can be visited by the client program linked
with a special library provided by CAEN running within the same TCP/IP
network.

A Pi will be used to collect the data from the gas handling
system. The most critical parameter is the flow rate, which represents
the stability of circulation gas flow. The flow rate will be measured
by a flow-meter (model MF5008~\cite{AITOLY:MF5008}), which generates
analog signals. The signals will be converted to digital values by an
analogy-digital converter (ADC) of the type DAM8082~\cite{DAM8082},
and sent to the SCS using RS485 protocol. The pressures in the gas
pipelines and the weights of the quenching gas cylinders~(TMA and
Isobutane) of the system will be read out by using the same pressure
sensor~(model PRC-905).  All the values are collected in the same way
as that of the flow rate and displayed by digital instruments (model
WSATC8030423HLP~\cite{WSAT}).  A RS485 hub (model
UT-1208~\cite{UT-1208}) will be used to combine the inputs to reduce
the consuming of the USB ports.  Pressure inside the TPC and power of
the Agilent 305FS~\cite{Agilent:305FS} vacuum pump are key parameters
during the pumping process. They represent the status and progress of
the process. The pump provides these values to the client program
running on the SCS server via the serial to USB cable using the RS232
protocol.

The DAQ system of PandaX-III is developed with the open-source
framework of Midas~\cite{Midas}. The number of trigger rates and
events are essential to onsite shifters, thus will be monitored by the
SCS. These values are sent directly to the InfluxDB by the DAQ
system, using the HTTP protocol.
	
The emergency recovery system is composed of some gas pipelines in the
gas handling system, the liquid nitrogen tank, gas recovery cylinders
and the dewar.  The recovery cylinder will be immersed in liquid
nitrogen, which is stored in the dewar as normal and will be replaced
regularly.  Once the xenon leakage is confirmed, the gas will be
recycled into the recovery cylinder.  The weight and the pressure of
the cylinder, and the level of liquid nitrogen are two key parameters
to decide whether the system could work.

The temperature and humidity inside the clean room, in which the
experiment runs, will be measured by a temperature and humidity sensor
(model AS109, produced by the ASAIR Inc.~\cite{AS109}). A Pi will be
deployed to collect the data from the sensor via the serial to USB
cable using the RS485 protocol. The air conditioner control system of
the cleanroom will generate CSV files to be read by the client program
running on the server.

The input voltage of the UPS, which is the most important indicator of
the power input, will be collected by the client program running on a
Pi via the USB cable with RS232 protocol. The output voltage and the
utility frequency, will be collected in the same way.

\section{Demonstrator of the Slow Control System}
\label{demonstrator}
\begin{figure}[htbp]
	\renewcommand{\figurename}{Fig.}
	\centering
	\includegraphics[width=0.9\linewidth]{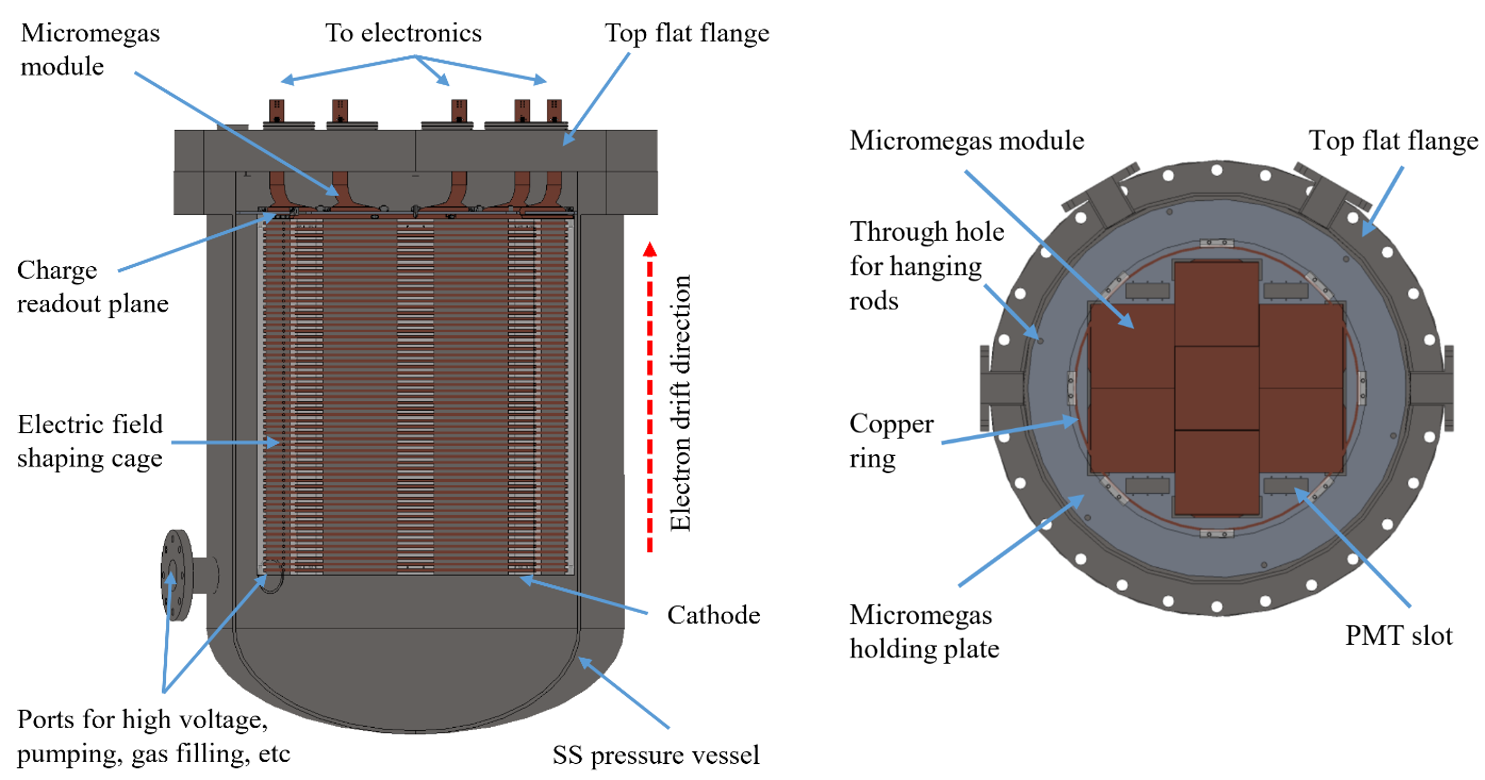}
	\caption{An illustration of the TPC~\cite{Lin:2018mpd}:
          (Left) cutaway drawing of the TPC and the pressure vessel
          with main components highlighted.  (Right) from the bottom
          up to highlight the charge readout plane with Micromegas
          modules.}
	\label{prototype}
\end{figure}

\begin{figure}[htbp]
	\renewcommand{\figurename}{Fig.}
	\centering
	\includegraphics[width=0.9\linewidth]{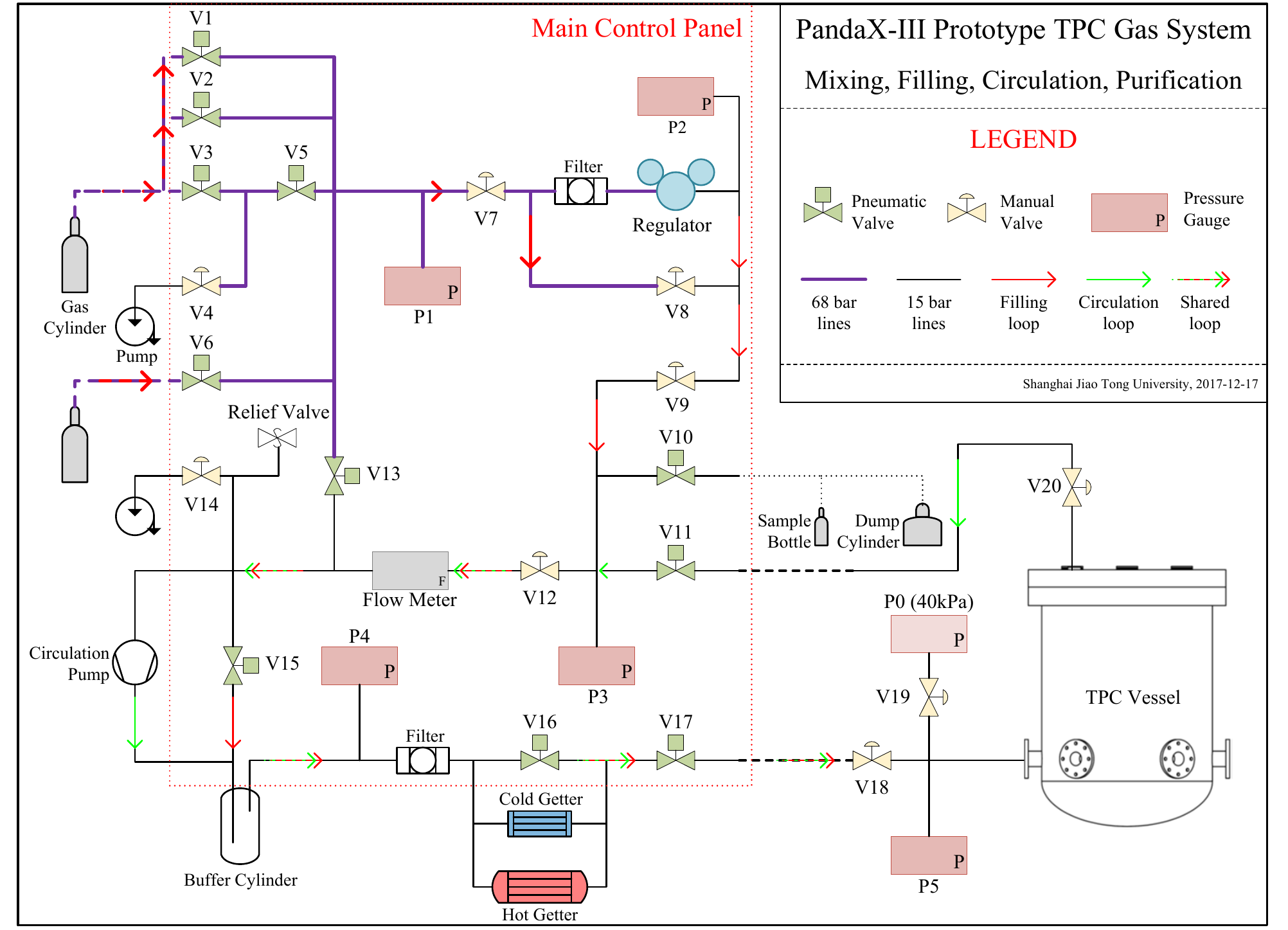}
	\caption{Schematic drawing for the gas handling
          system~\cite{Lin:2018mpd}.  Red arrows denote the filling
          loop. Green arrows denote the circulation loop.  Dotted red
          and green arrows are for the shared loop. Two getters are
          used for circulation and purification. All the key
          parameters in the system are monitored.}
	\label{gas system}
\end{figure}

A prototype detector is built to study the performance of
high-pressure xenon TPC for neutrinoless double beta decay search.
The detailed description of the components of the detector and its
subsystems is given in Ref.~\cite{Lin:2018mpd}.  As shown in
Fig.~\ref{prototype}, the detector vessel has an inner volume of about
600~L with a design operating pressure of 15~bar.  The active volume
in side the TPC is about 270~L, with 78~cm drift distance and a
readout plane composed of 7 Micromegas modules.  A gas handling system
(Fig.~\ref{gas system}) is used for the gas filling, circulation,
mixing, purification, and recovery of the prototype TPC. We set up a SCS demonstrator
(P-SCS) to monitor the key parameters. The pressures in the gas system and the TPC, and the
environment temperature/humidity were monitored for long time. The
high voltage and current for all the Micromegas and the cathode were
monitored during the data taking. Long term operation of the P-SCS had been carried on to test
its performance, and to optimize the design of the SCS as well.

\subsection{System set up}
The modules of data storage, data visualization, anomaly detection and
alarm dispatcher are running on a Dell R330 server, which is equipped
with a quad core Intel Xeon E5-2403v2 CPU, 8~GB of memory and a hard
drive with 1~TB capacity. The operating system is Ubuntu server 18.04
LTS. The same versions of InfluxDB, Chronograf and Kapacitor mentioned
in Section.~\ref{System design} are used. The default ports of 8086
(InfluxDB), 8088 (Chronograf) are not changed during the operation.
The alarm dispatcher uses the port of 3000, and is configured to send
the alarms to all the local members of the PandaX-III
collaboration. Parameters from the gas handling system, the UPS and
environment are collected by programs on a Pi, and other parameters
are collected by programs running on the server. The main parameters
monitored by P-SCS is given in Table~\ref{p-parameter}.
\begin{table}[htb]
  \centering
  \begin{tabular}{| c | c | c |}
    \hline Subsystems & Critical parameters & Data source\\
    \hline Prototype TPC & voltage, current & Matsusada AU\\
    \hline Micromegas & voltage, current & CAEN N1471, Iseg NHR4220\\
    \hline \multirow{2}*{Gas handling system} & pressure, pump power & Agilent 305FS\\
    \cline{2-3} & flow rate, pressures & MF5008\\
    \hline UPS & voltage, utility frequency & SANTAK Castle C3K\\
    \hline Environment & temperature, humidity & AS109 \\
    \hline
  \end{tabular}
  \caption{Overview of the parameters monitored by the P-SCS, together
    with the data sources.}
   \label{p-parameter}
\end{table}

\subsection{System operation}
The core modules of the P-SCS have been set up for about one year. The
client programs for different subsystems of the prototype detector
have been integrated when the corresponding systems are ready. For the
TPC and the Micromegas detector, the parameters are monitored only
during the data taking. Each run of data taking lasts for
hours. Details of the operation of the P-SCS are given below.

\begin{figure}[htbp]
	\renewcommand{\figurename}{Fig.}
	\centering
	\includegraphics[width=0.9\linewidth]{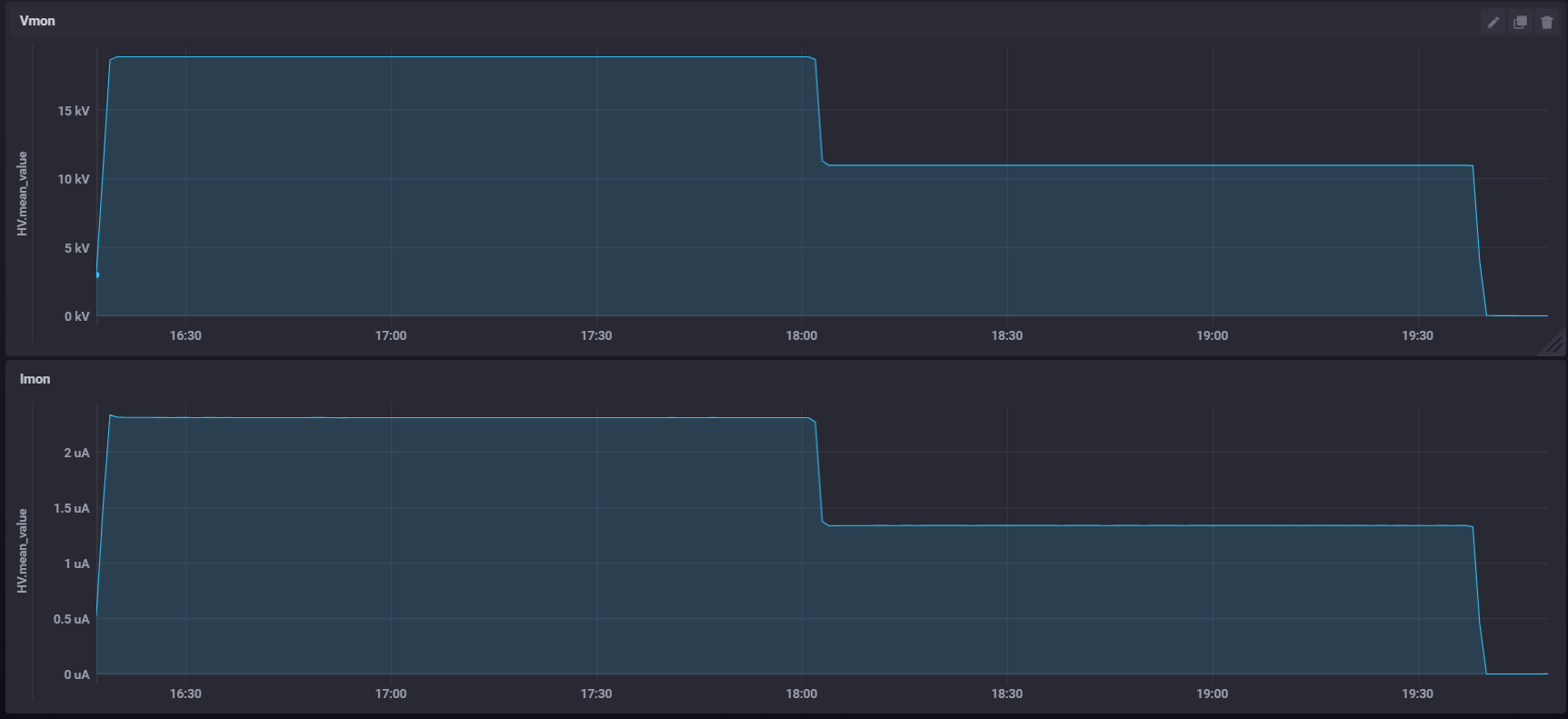}
	\caption{Example of the voltage~(top) and current~(bottom) from the Matsusada AU
		HV power supplies during a data taking run, collected by the
		P-SCS.}
	\label{HV}
\end{figure}

\begin{figure}[htbp]
	\renewcommand{\figurename}{Fig.}
	\centering
	\subfigure[Alert message on the alert web]{
		\label{fig:subfig:webalert}
		\includegraphics[width=0.6\linewidth]{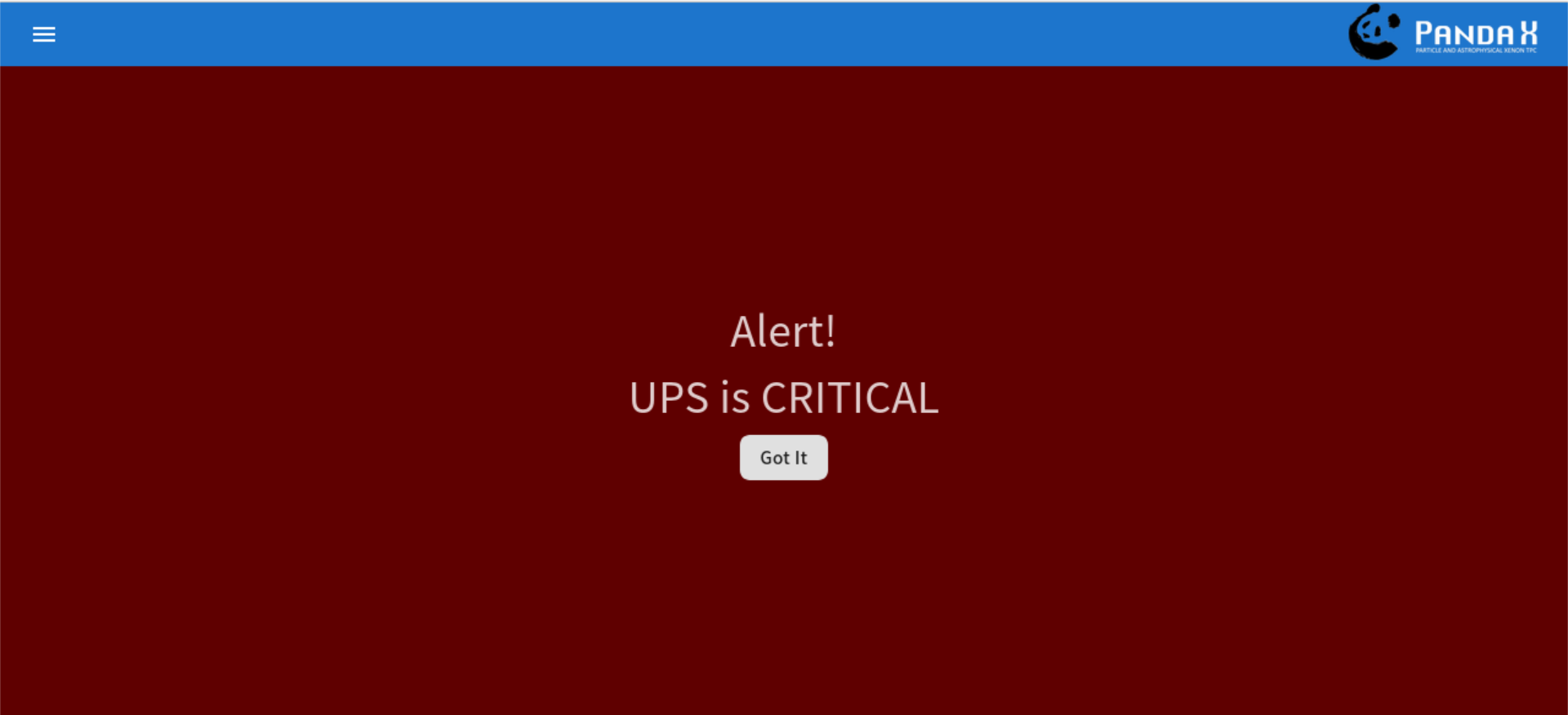}	
	}
	\subfigure[Alert message on Enterprise WeChat]{
		\label{fig:subfig:wechat}
		\includegraphics[width=0.35\linewidth]{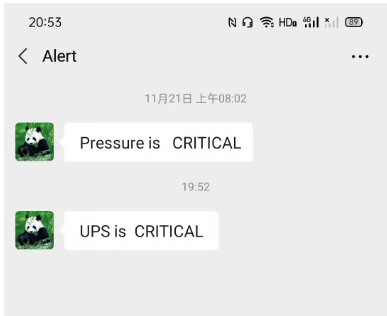}
	}
	\caption{An alert message was generated by the accident of input power cut of the UPS. 
	The message is displayed on the web~(a) and delivered to users by the Enterprise WeChat~(b).}
	\label{alert}
\end{figure}

\begin{itemize}
\item {\bf Prototype TPC}: The voltage and the current of the cathode
  are monitored with an interval of 5~s. The interval enables us to
  obtain a precise picture of the evolution of HV during data
  taking. An example of the collected data in one data taking run,
  during which we examined the detector performance with 2 different
  HVs of the cathode, is given in Fig.~\ref{HV}. One can spot the
  rising of voltage and current at the beginning, and the dropping of
  them during the run clearly.

\item {\bf Micromegas}: The voltage and current of the Micromegas
  detector are monitored with an interval of 1~s. In comparison with
  that of the TPC, the interval is smaller, which is important for the
  monitoring of the sparking of the Micromegas in time. The parameters
  of one Micromegas during a complete data taking run are given in
  Fig.~\ref{N1471}. We increased the voltage from 0 to 350 V for safe
  operation of the Micromegas at a pace of 3~V/s, accompanied by a
  small current for charging. Then the voltage was increased to 390 V
  for data taking. We observed 9 times of sparking, during which the
  current increased dramatically. The onsite shifter stopped the run
  and turned down the voltage due to the large number of sparking with
  current over the threshold of 0.3~$\mu$A.
  \begin{figure}[htb]
    \renewcommand{\figurename}{Fig.}
    \centering
    \includegraphics[width=0.9\linewidth]{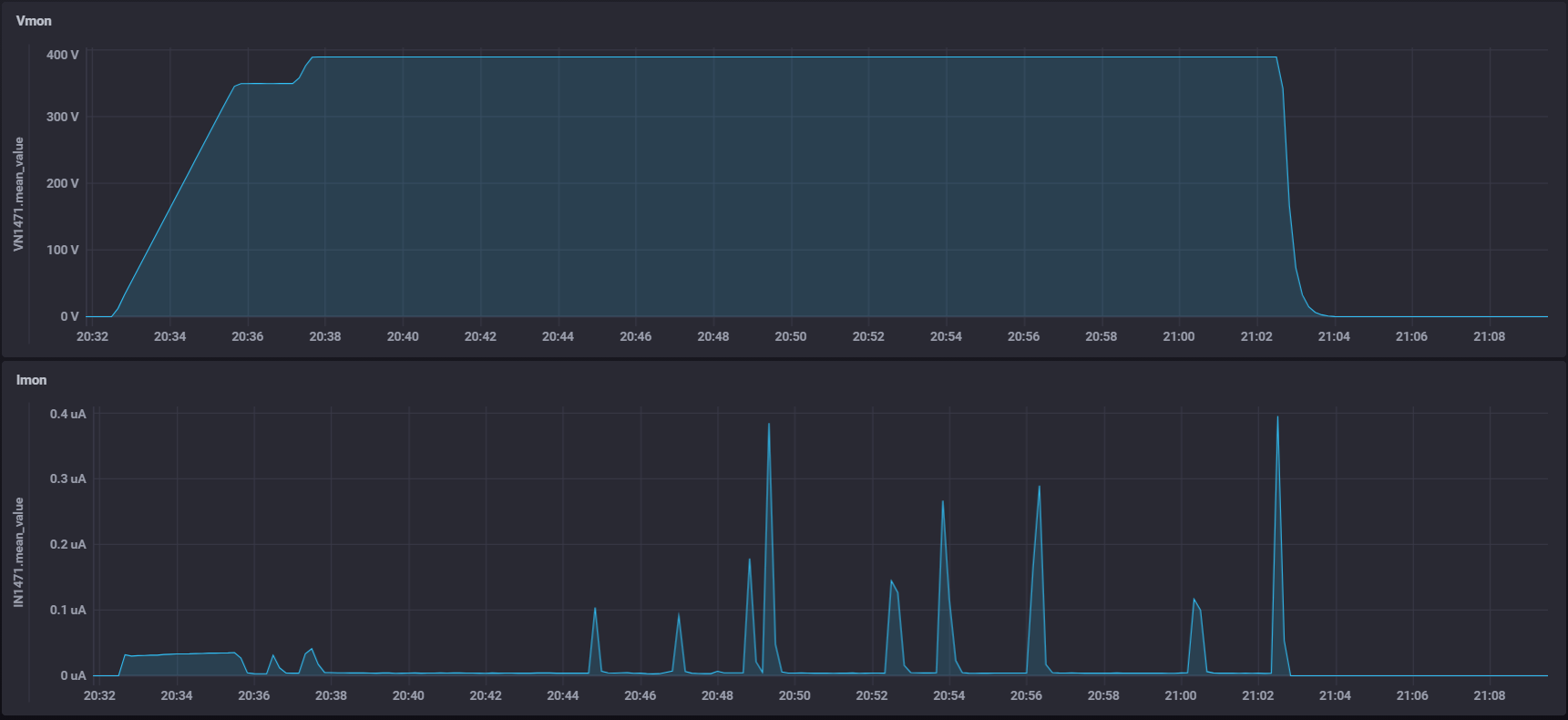}
    \caption{Voltage~(top) and current~(bottom) of one Micromegas during a data
      taking run.}
    \label{N1471}
  \end{figure}

\item {\bf Gas handling system}: The pressures within the gas
  pipelines and weight of the quenching gas cylinders connected in the
  system are monitored at an interval of 5~s.  Fig.~\ref{gas} shows
  the evolution of the pressure at both ends of the circulating pump
  during one of the gas circulation.  A positive difference between
  the pressures should be kept during a normal circulation. No
  flow-meter was integrated in the pipelines, because the mixed TMA
  may be corrosive to the rubber O-ring of the flow-meter. In the
  PandaX-III experiment, flow-meters will be added back by replacing
  current O-rings with those made of PTFE.
\begin{figure}[htb]
	\renewcommand{\figurename}{Fig.}
	\centering
	\includegraphics[width=0.9\linewidth]{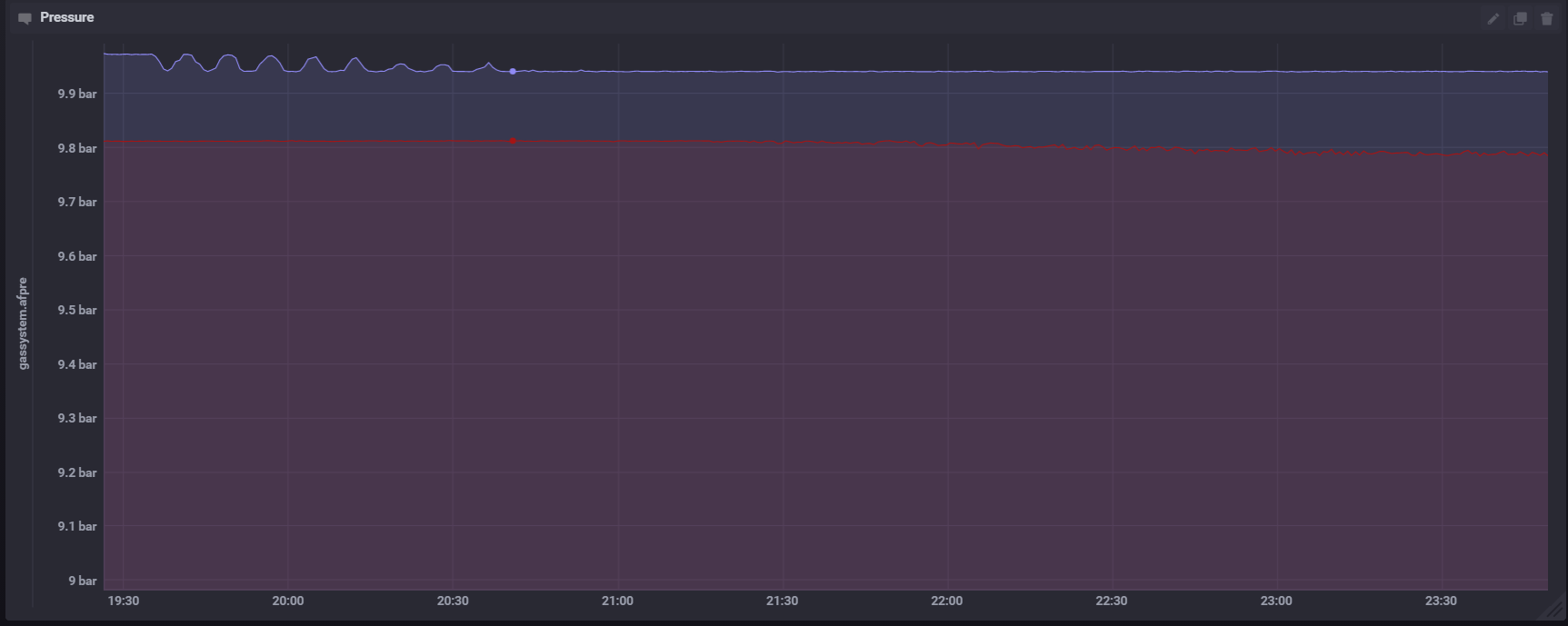}
	\caption{Pressures at both ends of the pump during one of
          the gas circulation, 
          P3 and P4 in Fig.~\ref{gas system}.
          The blue curve represents the pressure at the front end of the circulation pump, the red one is the evolution of the pressure at the back end.}
	\label{gas}
\end{figure}

\item {\bf Environment}: The temperature and humidity of the working
  space of the detector are monitored with an interval of 5~s since
  January of 2020.  Fig.~\ref{AS109} shows the evolution of these
  parameters till the writing of the report. A break of about 5 days
  on the picture was caused by the temporarily removing of the Pi,
  which served as the data collector. Monitoring of other parameters
  was not affected by this event. The temperature and humidity changed
  according to the change of seasons. No failure of the Pi happened
  during the long time data taking.
  \begin{figure}[htb]
    \renewcommand{\figurename}{Fig.}
    \centering
    \includegraphics[width=0.9\linewidth]{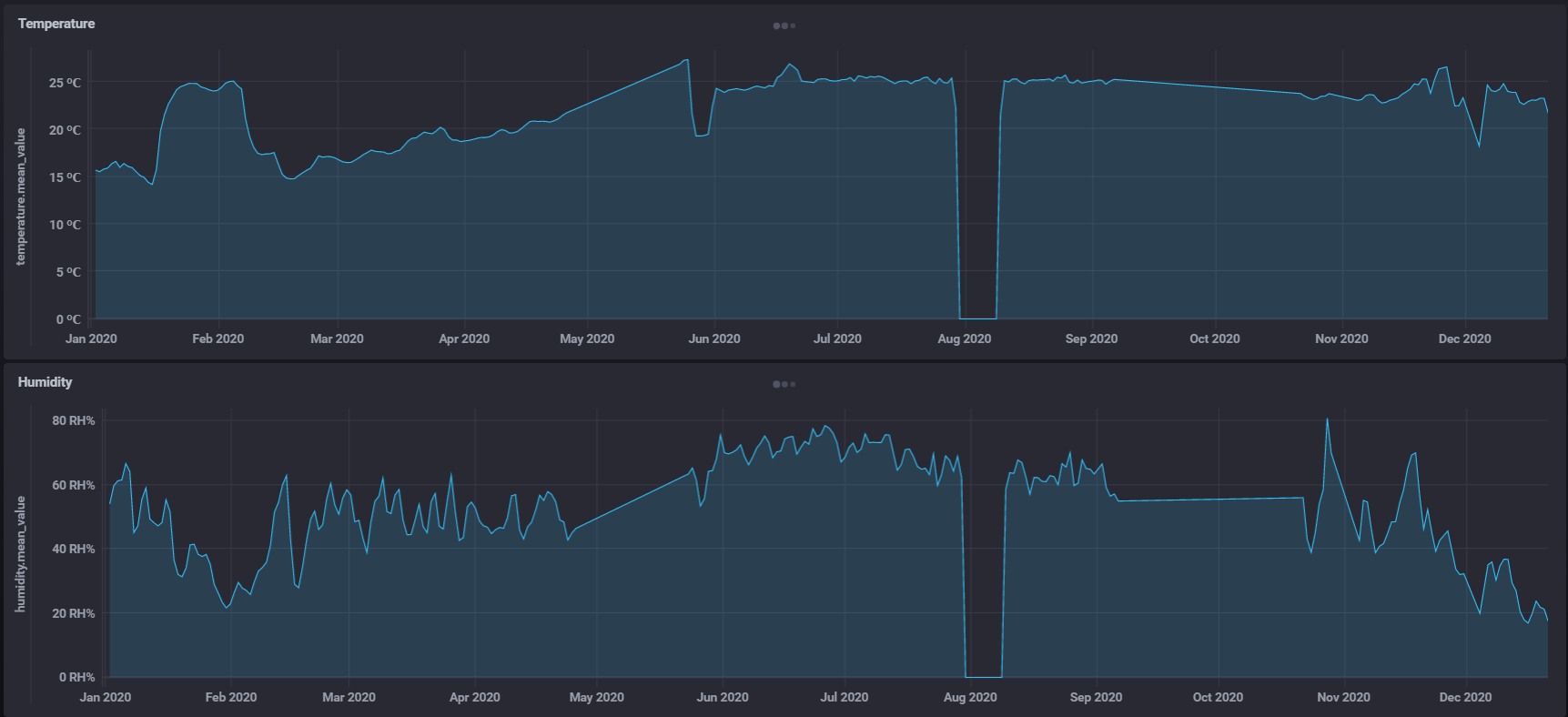}
    \caption{Evolution of the temperature~(top) and humidity~(bottom) of the clean room
      collected by a Pi from Jan., 2020.}
    \label{AS109}
  \end{figure}

\item {\bf UPS}: The parameters of the UPS have been monitored since
  November, 2020, at an interval of 5~s. Evolution of the input
  voltage is given in Fig.~\ref{ups}. Fluctuation of the voltage
  implies the quality of the external power supply. Due to the circuit
  maintenance, the power was cut for two hours on December 22, 2020
  and the alert dispatching module displayed the alert message on the
  alert web~(Fig.~\ref{fig:subfig:webalert}).  The local member
  received the alert message through Enterprise
  WeChat~(Fig.~\ref{fig:subfig:wechat}) at nearly the same time.
  
  \begin{figure}[htb]
    \renewcommand{\figurename}{Fig.}
    \centering
    \includegraphics[width=0.9\linewidth]{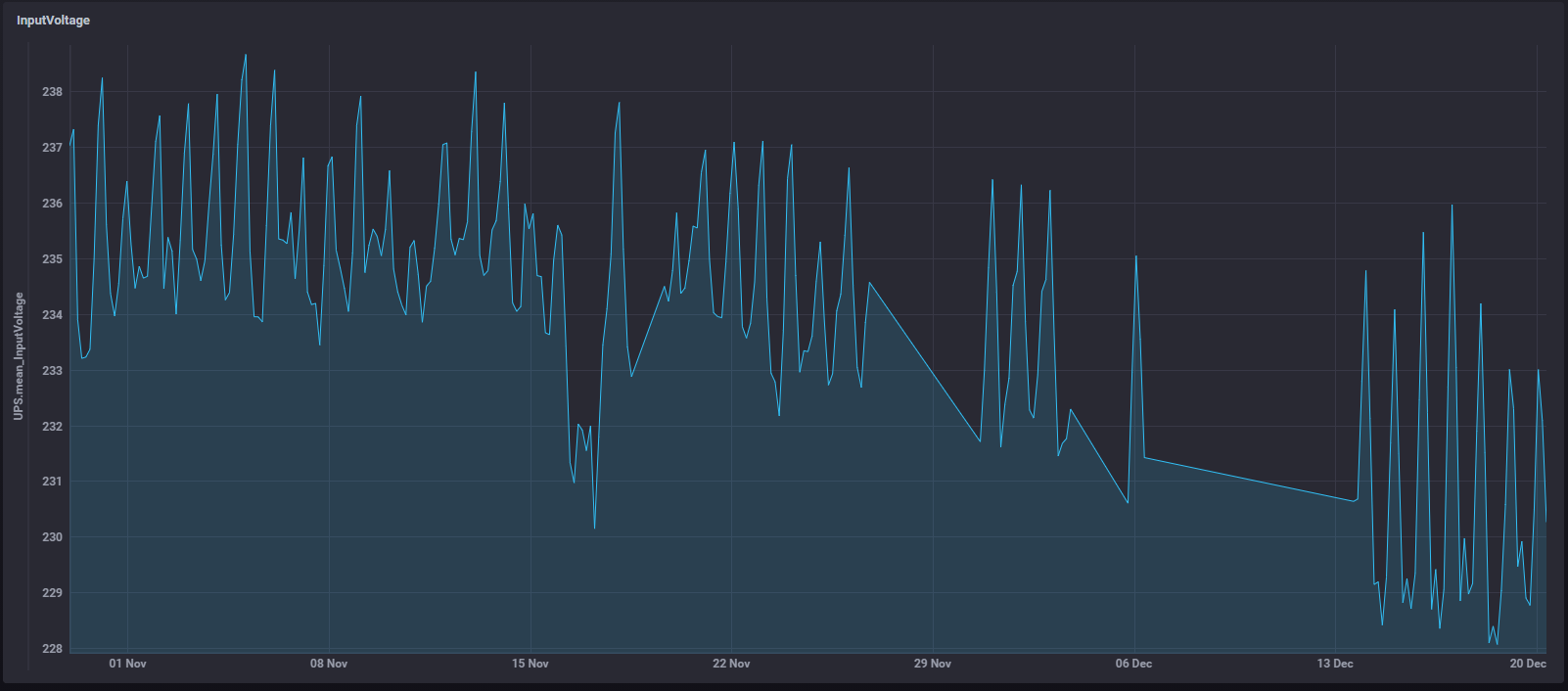}
    \caption{Input voltage of the UPS from Nov., 2020.}
    \label{ups}
  \end{figure}
\end{itemize} 

The P-SCS is running since January 1, 2020. The server, which contains
most of the modules, works stably. It has been shut down manually for
2 times due to the power cut, but no failure was observed during the
normal operation. The CPU and memory consumption of the software are
kept at a stable level during normal operation. Increment of the
consumption has been observed at the time when a user tried to query
data within a large time span, but still at a safe level. This
indicates that larger memory and powerful CPUs are required by the SCS
server of the PandaX-III experiment.  Besides, the Pi used to collect
data also shows good stability. To keep the system monitored without
interrupt, all the components of the SCS should be driven by the
UPS. Though the Pi used in the P-SCS behaves stable, more backups are
required to response to the possible failure. We also checked the disk
consumption of the system. After one year of the operation, with all
the parameters mentioned above, only 4 GB of disk space is consumed by
the data. Based on this value, by considering a complete SCS for the
PandaX-III experiment, we estimate that about 100 GB of disk space
will be consumed each year. The experience from the operation of the
P-SCS provides important reference to the future PandaX-III SCS.

\section{Conclusion}
\label{conclusion}
A SCS with five loosely coupled modules for the PandaX-III experiment
is designed. A demonstrator of the SCS is set up to monitor the
critical parameters of the prototype detector. All the modules have
been tested. The built system behaves stable during the operation of
one year. In the near future, a complete SCS will be set up together
with the PandaX-III experiment. The system will provide all the
critical information of the facilities to the experimentalists, acting
as an essential safeguard.

\section*{Acknowledgements}
This work is supported by the grant from the Ministry of Science and Technology of China~(No.2016YFA0400302) 
and the grants from National Natural Sciences Foundation of China~(No.11775142 and No.11905127). 
We thank the support from the Key Laboratory for Particle Physics, Astrophysics and Cosmology, Ministry of Education. 
This work is supported in part by the Chinese Academy of Sciences Center for Excellence in Particle Physics~(CCEPP).
We thank Chen Cheng in PandaX-4T group to give us some suggestions in the design and implementation of the whole system.


\bibliography{mybibfile}

\begin{thebibliography}{10}

\bibitem{Chen:2016qcd}
X.~Chen et~al.
\newblock {PandaX-III: Searching for neutrinoless double beta decay with high
  pressure$^{136}$Xe gas time projection chambers}.
\newblock {\em Sci. China Phys. Mech. Astron.}, 60(6):061011, 2017.

\bibitem{Wang:2020owr}
S.~Wang.
\newblock {The TPC detector of PandaX-III Neutrinoless Double Beta Decay
  experiment}.
\newblock {\em JINST}, 15(03):C03052, 2020.

\bibitem{Giomataris:1998rc}
Y.~Giomataris.
\newblock {Development and prospects of the new gaseous detector `Micromegas'}.
\newblock {\em Nucl. Instrum. Meth. A}, 419:239--250, 1998.

\bibitem{Wang:2020csx}
S.~Wang.
\newblock {PandaX-III high pressure xenon TPC for Neutrinoless Double Beta
  Decay search}.
\newblock {\em Nucl. Instrum. Meth. A}, 958:162439, 2020.

\bibitem{Ji:2019cwn}
P.~Ji et~al.
\newblock {A low-cost slow control system for the PandaX-4T experiment}.
\newblock {\em Radiat. Detect. Technol. Methods}, 3(3):53, 2019.

\bibitem{InfluxDB}
InfluxDB, Chronograf, Kapacitor:
  \url{https://portal.influxdata.com/downloads/}.

\bibitem{c-sharp}
\emph{C\#}: https://docs.microsoft.com/en-us/dotnet/csharp/.

\bibitem{Python}
Python: \url{https://www.python.org/}.

\bibitem{systemd}
Systemd: \url{https://systemd.io/}.

\bibitem{RaspberryPi:4B}
RaspberryPi-4B:
  \url{https://www.raspberrypi.org/products/raspberry-pi-4-model-b/}.

\bibitem{RS485}
RS485: \url{https://www.definitions.net/definition/RS-485}.

\bibitem{RS232}
RS232: \url{https://www.definitions.net/definition/RS-232}.

\bibitem{falcon}
Falcon: \url{https://falconframework.org/}.

\bibitem{Gunicorn}
Gunicorn: \url{https://gunicorn.org/}.

\bibitem{WeChat}
WeChat: \url{https://work.weixin.qq.com}.

\bibitem{reactjs}
Reactjs: \url{https://reactjs.org/}.

\bibitem{Matsusada:AU}
Matsusada AU power supply:
  \url{https://www.matsusada.com/product/hvps1/rack/au/}.

\bibitem{prc}
PRC-905: \url{http://www.runchentec.com/product_detail/6.html}.

\bibitem{Zhu:2018oxf}
S.~Liu et~al.
\newblock {Development of the Front-End Electronics for PandaX-III Prototype
  TPC}.
\newblock {\em IEEE Trans. Nucl. Sci.}, 66(7):1123--1129, 2019.

\bibitem{CAEN}
SY2527 Universal Multichannel Power Supply System and N1471 HV Power Supply
  Module (USB): \url{https://www.caen.it/download/}.

\bibitem{AITOLY:MF5008}
MF5008: \url{https://www.aitoly.com}.

\bibitem{DAM8082}
DAM8082: \url{http://www.c-control.cn/mobile/}.

\bibitem{WSAT}
WSATC8030423HLP: \url{http://www.runchentec.com/product_detail/17.html}.

\bibitem{UT-1208}
UT-1208: \url{http://www.szutek.com/pro_view-98.html}.

\bibitem{Agilent:305FS}
TwisTorr 305 FS: \url{https://www.agilent.com/search/?Ntt=305FS}.

\bibitem{Midas}
Midas: \url{https://midas.triumf.ca}.

\bibitem{AS109}
Asair AS109: \url{http://www.aosong.com/products-52.html}.

\bibitem{Lin:2018mpd}
H.~Lin et~al.
\newblock {Design and commissioning of a 600 L Time Projection Chamber with
  Microbulk Micromegas}.
\newblock {\em JINST}, 13(06):P06012, 2018.

\end{thebibliography}
\bibliographystyle{unsrt}
\appendix
\section{The client program}
\subsection{Set up of the client program}
\label{On-Shot}
In the ``on-shot'' mode, the client program will be executed only
once. The key function is given below.

\begin{lstlisting}[language=Python,showstringspaces=false]
def read_and_post(self, device):
  try:
  #check the status of device serial port
   if not device.open()==False: 
	 for s in device.sensors:
	     #read data from device
	     v = s.read()   
	     if v is not None and len(v) != 0:
			logging.debug("%s %s", s.name, str(v))
			#send data to the database
			self.post(s, v)  
		 else:
			logging.info("No data.")
	#Close the port to finish the reading process
	device.close()  
  except Exception as err:
         logging.exception(str(err))
         device.close()
\end{lstlisting}

\subsection{Set up of the timer}
\label{Timer}
The example files to setup the systemd Timer service for the monitoring of the TPC is given below.

The ``Serivce'' file:
\begin{lstlisting}[language=Python]
[Unit]
Wants=user@.service
[Service]
#start the client program at specified location
ExecStart = python3 /home/pi/p3sc/reader.py -c /home/pi/p3sc/tpc.yaml 
[Install]
WantedBy=default.target
\end{lstlisting}

The ``Timer'' file:
\begin{lstlisting}[language=Python]
[Unit]
Description=p3sc timer
[Timer]
# Time to wait after enable this unit
OnActiveSec=0
# Time between running each consecutive time
OnUnitActiveSec=5
#Specify the object of timer
Unit=p3sc-tpc.service
#Setting the trigger precision of timer
AccuracySec=1
[Install]
WantedBy=default.target
\end{lstlisting}

\subsection{Set up of data buffer}
\label{back up} 
The data buffer is implemented as a queue of fixed length. Once the
client program disconnects with InfluxDB, the readout values will be
stored in this queue, and written to a local text file when the queue is
full. When the connection is back, the values in the queue and the
text file will be posted to InfluxDB again. The related code is given below:
\begin{lstlisting}[language=Python,label=queue,showstringspaces=false]
def post():
  url='http://192.168.3.41:8086' + '/write?db=' + 'temp'
  #check the status of queue 
  if not buffer_queue.empty():
     #get data from queue
     data = buffer_queue.get()
     try:
       #try to send data to the database 
       requests.post(url, data = data)
     except Exception as e:
       print("===> connect lost! <===")
       if buffer_queue.full():
         with lock:
           #the data will be stored in the text file if the
             #queue is full
           with open("buffer.txt", 'a') as buffer_file:
             buffer_file.write(data)
           print("Overflow! Data has been written to buffer!")
           pass
       else:
          #If the queue is not full, the data, which fails 
           #to be posted to the database, will be stored 
           #in it.
          buffer_queue.put(data)
       print("WARNING: Failed to post the package to server!")
  else:
     print("===> successed!")

\end{lstlisting}

\begin{figure}[htbp]
	\renewcommand{\figurename}{Fig.}
	\centering
	\includegraphics[width=0.8\linewidth]{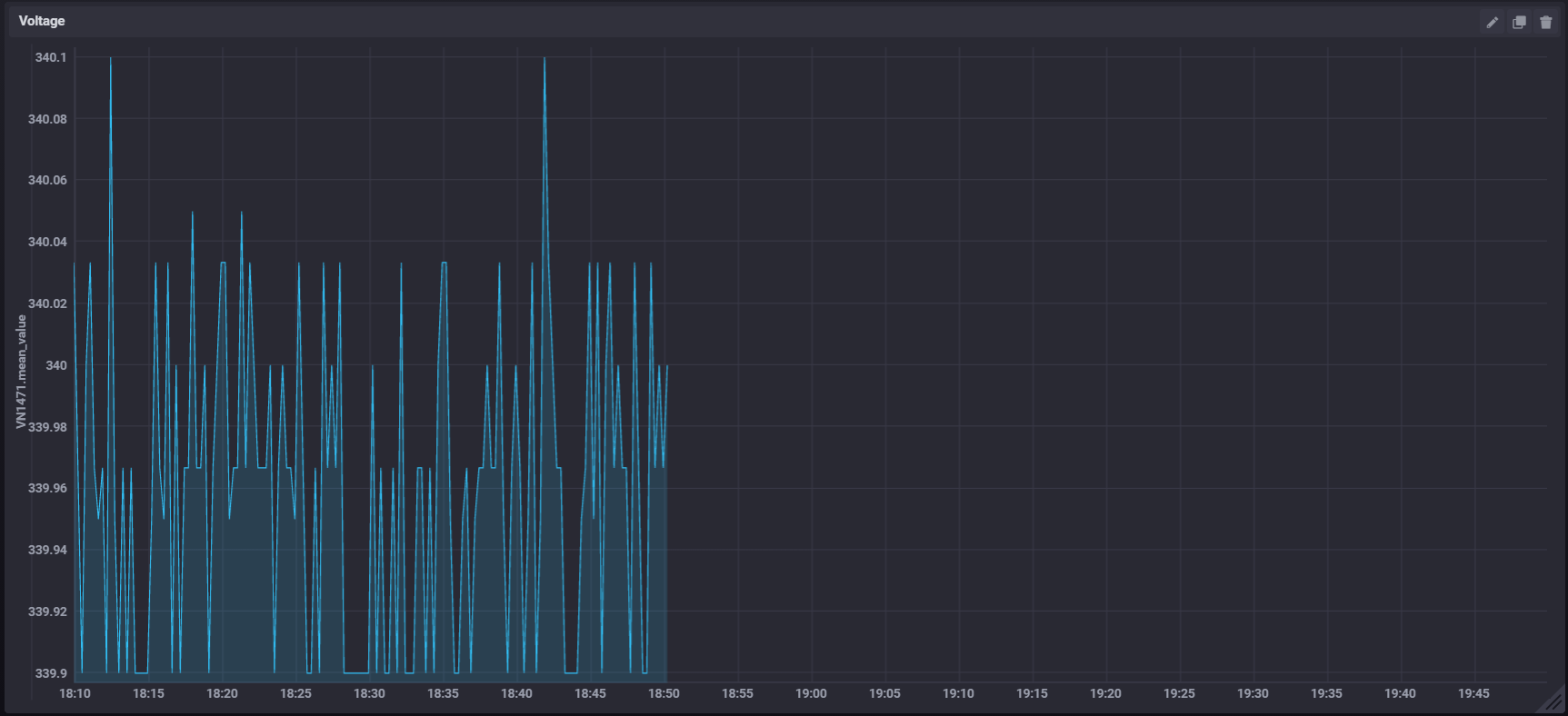}
	\includegraphics[width=0.8\linewidth]{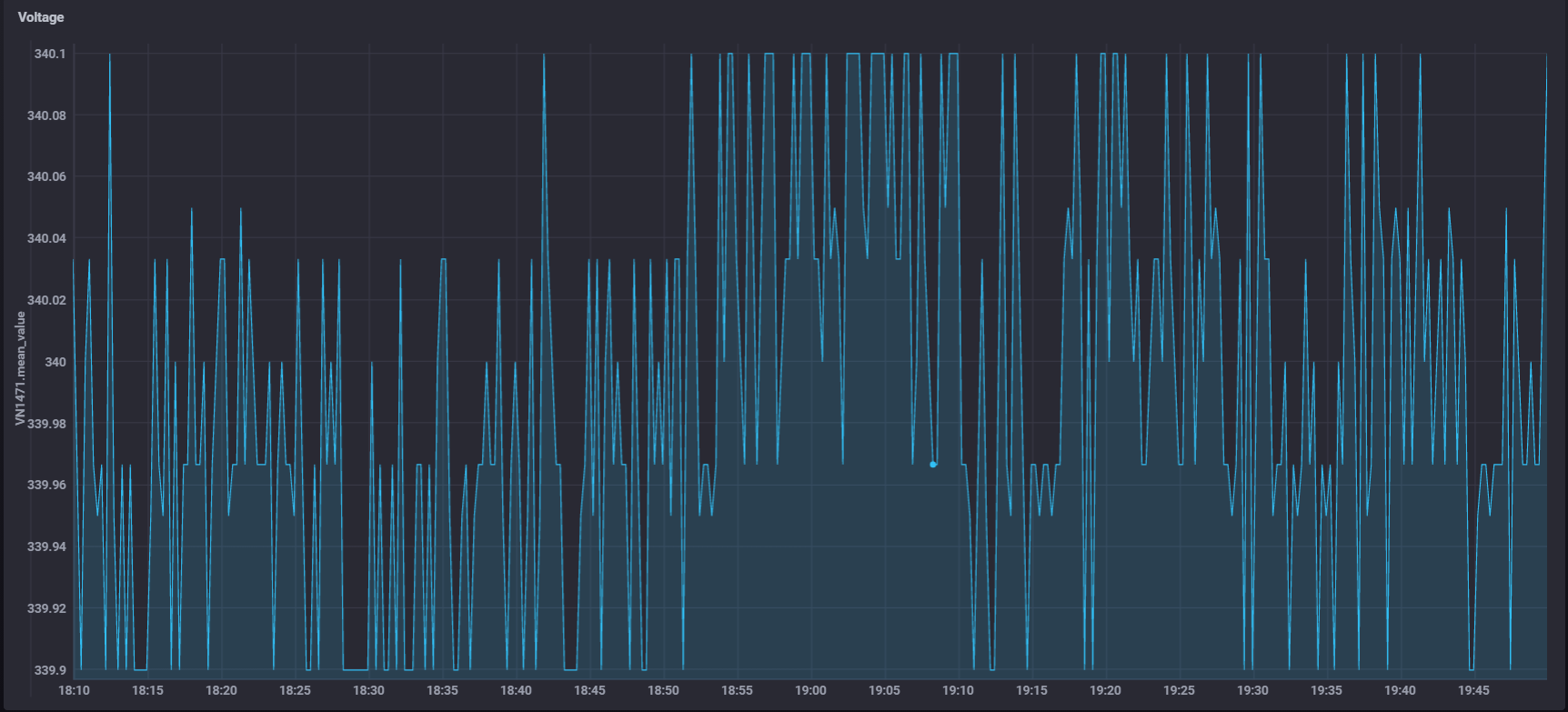}
	\caption{Voltage of one Micromegas during a data taking run
          for network interruption test. The data is unavailable in
          the database when the client box is disconnected from the
          network~(Top), and restored when the box is reconnected to
          the network (Bottom).}
	\label{voltageloss}
\end{figure}

Fig.~\ref{voltageloss} shows an example of voltage monitoring of one
Micromegas, the network was disconnected for about one hour. Then all
the buffered information was restored when the network is back.

\end{document}